\documentclass[twocolumn]{aastex62}
\usepackage{xcolor} % Required for specifying custom colors
\usepackage{booktabs}
\usepackage{cleveref}
\usepackage{longtable}
\usepackage[normalem]{ulem}

\definecolor{dartmouthgreen}{rgb}{0.05, 0.5, 0.06}
\newcommand{\ch}[1]{{\color{black} #1}}

\newcommand{\DM}{{\ensuremath{\rm DM}}}
\newcommand{\sigmatoa}{{\ensuremath{\sigma_{\rm TOA}}}}

\newcommand{\U}{\ensuremath{T}}
\newcommand{\dmunit}{{\ensuremath{{\rm pc}\,{\rm cm}^{-3}}}}

\def\be{\begin{equation}}
\def\ee{\end{equation}}
\def\ba{\begin{eqnarray}}
\def\ea{\end{eqnarray}}

\makeatletter
\define@key{Gin}{resolution}{\pdfimageresolution=#1\relax}
\makeatother
\submitjournal{ApJ}

\shorttitle{}
\shortauthors{Jones et al.}

\begin{document}

\title{Evaluating Low-Frequency Pulsar Observations to Monitor Dispersion with the Giant Metrewave Radio Telescope }

\correspondingauthor{Megan~L.~Jones}
\email{megan.jones@nanograv.org}

\author[0000-0001-6607-3710]{M.~L.~Jones}
\affil{Center for Gravitation, Cosmology and Astrophysics, Department of Physics, University of Wisconsin-Milwaukee, P.O. Box 413, Milwaukee, WI 53201}
\affil{Department of Physics and Astronomy, West Virginia University, Morgantown, WV 26506, USA}
\affil{Center for Gravitational Waves and Cosmology, West Virginia University, Chestnut Ridge Research Building, Morgantown, WV 26505, USA}

\author[0000-0001-7697-7422]{M.~A.~McLaughlin}
\affil{Department of Physics and Astronomy, West Virginia University, Morgantown, WV 26506, USA}
\affil{Center for Gravitational Waves and Cosmology, West Virginia University, Chestnut Ridge Research Building, Morgantown, WV 26505, USA}

\author[0000-0002-2892-8025]{J.~Roy}
\affil{National Centre for Radio Astrophysics, Tata Institute of Fundamental Research, Pune 411 007, India}

\author[0000-0003-0721-651X]{M.~T.~Lam}
\affil{School of Physics and Astronomy, Rochester Institute of Technology, Rochester, NY 14623, USA}
\affil{Laboratory for Multiwavelength Astrophysics, Rochester Institute of Technology, Rochester, NY 14623, USA}

\author[0000-0002-4049-1882]{J.~M.~Cordes}
\affil{Cornell Center for Astrophysics and Planetary Science and Department of Astronomy, Cornell University, Ithaca, NY 14853, USA}

\author[0000-0001-6295-2881]{D.~L.~Kaplan}
\affil{Center for Gravitation, Cosmology and Astrophysics, Department of Physics, University of Wisconsin-Milwaukee, P.O. Box 413, Milwaukee, WI 53201}

\author[0000-0002-6287-6900]{B.~Bhattacharyya}
\affil{National Centre for Radio Astrophysics, Tata Institute of Fundamental Research, Pune 411 007, India}

\author[0000-0002-2034-2986]{L.~Levin}
\affil{Jodrell Bank Centre for Astrophysics, School of Physics and Astronomy, University of Manchester, Manchester M13 9PL, UK}

\begin{abstract}
The North American Nanohertz Observatory for Gravitational Waves (NANOGrav) project has the primary goal of detecting and characterizing low-frequency gravitational waves through high-precision pulsar timing.
The mitigation of interstellar effects is crucial to achieve the necessary precision for gravitational wave detection. 
Effects like dispersion and scattering are stronger at lower observing frequencies, with the variation of these quantities over week--month timescales requiring high-cadence multi-frequency observations for pulsar timing projects. 
In this work, we utilize the dual-frequency observing capability of the Giant Metrewave Radio Telescope (GMRT) and evaluate the potential decrease in dispersion measure (DM) uncertainties when combined with existing pulsar timing array data. 
We present the timing analysis for four millisecond pulsars observed with the GMRT simultaneously at 322 and 607\,MHz, and compare the DM measurements with those obtained through NANOGrav
observations with the Green Bank Telescope (GBT) and Arecibo Observatory at 1400 to 2300\,MHz frequencies. 
Measured DM values with the GMRT and NANOGrav program show significant offsets for some pulsars, which could be caused by pulse profile evolution between the two frequency bands. 
In comparison to the predicted DM uncertainties when incorporating these low-frequency data into the NANOGrav dataset, we find that higher-precision GMRT data is necessary to provide improved DM measurements. 
Through the detection and analysis of pulse profile baseline ripple in data on test pulsar B1929+10, we find that, while not important for these data, it may be relevant for other timing datasets. 
We discuss the possible advantages and challenges of incorporating GMRT data into NANOGrav and International Pulsar Timing Array datasets. 
\end{abstract}


\section{Introduction} \label{sec:intro}

{Gravitational waves (GWs) offer a new window through which to study the Universe, with the first direct GW detection in 2016 \citep{ligo}.}
A low-frequency ($\sim$nHz) detection in the pulsar timing array (PTA) portion of the GW spectrum would {provide information about} sources (e.g., the GW background due to supermassive black hole binaries, supermassive black hole mergers, among others) that are not visible to other GW experiments sensitive to higher frequency GWs \citep[e.g.,][]{ses13,las16,arz20,buc20,bla20}. 
In order to detect GWs using PTAs,
timing models for each pulsar must first be constructed by accounting for all known effects on the pulse times of arrival (TOAs) in order to minimize the differences between the measured and model-predicted TOAs (i.e., timing residuals).
The detection of GWs using pulsars requires high-precision timing with TOA accuracy less than $\sim$microseconds \citep[e.g.,][]{dem13,arz15}. Therefore all timing fluctuations, both intrinsic (e.g., binary motion) and extrinsic (e.g., interstellar plasma) to the pulsar, must be accounted for in the timing model.

There are currently three global PTA efforts focused on GW detection through pulsar timing {with decade-long data sets}: the North American Nanohertz Observatory for Gravitational Waves 
\citep[NANOGrav;][]{arz18}, the European Pulsar Timing Array \citep[EPTA;][]{des16}, and the Parkes Pulsar Timing Array \citep[PPTA;][]{ker20}. 
These three experiments {along with the more recently formed Indian Pulsar Timing Array \citep[InPTA;][]{jos18}}  form the International Pulsar Timing Array \citep[IPTA;][]{per19}. IPTA data releases are comprised of data from all eight radio telescopes used by these three regional collaborations. As more pulsar observing instruments have come online, additional PTA efforts are developing \citep[e.g.,][]{jos18,hob19,bai20}. The inclusion of more instruments into the IPTA can aid in producing a more valuable dataset by increasing the number of MSPs, the number of TOAs, the sky coverage, and observing frequency coverage. \ch{{In addition, {new low-frequency ($<200$\,MHz)} instruments provide useful supporting measurements to PTA science, such as long-term DM and pulse broadening variations of both millisecond and canonical pulsars, that can be incorporated into IPTA analyses \citep[e.g.,][]{bha18,kir19,ban19,bil20}.}
}

The Giant Metrewave Radio Telescope (GMRT) consists of 30 antennas, each with a 45-meter diameter; the total collecting area of the GMRT is equivalent to a $\sim$250-m diameter single dish telescope. 
Using six feeds, the array can observe finite frequency bands with centers ranging from 150 to 1250\,MHz \citep{gup17}. By splitting the array, the GMRT is capable of executing simultaneous dual-frequency observations that can be complementary to NANOGrav observations by offering coverage at lower frequencies, where frequency-dependent timing fluctuations due to interstellar effects are more prominent {because of the inverse dependence on observing frequency}.

\subsection{Interstellar Medium Effects}

The dominant delay induced by the interstellar medium (ISM) in timing data is due to dispersion. As the radio pulse travels through the ISM, it encounters ionized plasma along the way. 
The dispersion measure (DM) is the integrated column density of free electrons along the line of sight (LOS) to a pulsar
\begin{equation}
	\rm{DM} = \int_{0}^{\it{d}} \it{n_e(l) dl}~,
	\label{eq:def}
\end{equation}
where $n_e$ is the free electron density {at position $l$ along the LOS of distance $d$} to the pulsar. 
DM therefore can be used to infer the distance to the pulsar by assuming a free electron density model for the Galaxy \citep[e.g.,][]{cor02,yao17}. 
The time delay due to dispersion is
\begin{equation}
t=K \frac{\DM}{\nu^2},
\end{equation}
where  $K$=4.15\,ms\,GHz$^2$\,pc$^{-1}$\,cm$^{3}$ is the dispersion constant {\citep{lor12}} and $\nu$ is the observing frequency. DM can be estimated by observing at multiple frequencies or at a single frequency over a wide bandwidth and comparing the respective time delays. 
Due to the changing LOS, {both} as the pulsar moves relative to the Earth {as well as dynamic processes in the ISM}, DM is not constant in time and is in reality DM($t$), requiring epoch-to-epoch monitoring and correction {\citep[e.g.,][]{isa77,jon17}}.

In addition to dispersion, interstellar scattering also causes a frequency-dependent time delay, {scaling as $\sim\nu^{-4}$ and therefore much more influential at lower observing frequencies.} 
As the pulse travels through the ISM, it will be scattered due to inhomogeneities in the ISM which cause multi-path propagation. These multiple ray paths introduce a delay in the TOA. 
Like DM, scattering delays are also time variable. 
Scattering cannot be corrected as easily as DM, but can be partially corrected using high-resolution dynamic spectra due to the similar phenomenological cause behind scattering and scintillation \citep{lev16}. 
Because of the covariances between fitting for DM and scattering,
some scattering effects will be absorbed by fitting for only DM. 
Thus low-frequency observations can be used to disentangle the scattering contributions from the DM {through the discrimination between the $\nu^{-2}$ and $\nu^{-4}$ variations}.

\subsection{This Work}
NANOGrav typically observes pulsars with a strategy designed to measure DM variations at the cost of observing time \citep[e.g.,][]{arz18}. 
Most MSPs are observed on a monthly cadence with a single telescope --- either the Arecibo Observatory or the Green Bank Telescope (GBT) --- using observations at two frequencies: a higher frequency (typically 1.4\,GHz or above) where the timing precision is often better (see \citealt{lam18}) and a lower frequency (typically 800\,MHz or below) to provide a long frequency lever arm to track DM variations. This doubles the required observing time per source beyond the minimum required to just measure the TOAs themselves. In this work we examine whether incorporating observations from a third telescope can help anchor the DM measurements {to the DM curve},  increase their precision, and provide a valuable boost in observing efficiency.

{Through this analysis, we help} evaluate low-frequency data obtained using the GMRT for ultimate inclusion into the IPTA. We present GMRT timing data for four MSPs that are also part of the NANOGrav 11-year dataset. 
We investigate the potential improvement in DM precision by incorporating this GMRT data with the NANOGrav 11-year data as a test case. 
We discuss the data acquisition and observing modes at the GMRT in \S\ref{sec:data}. 
We compare predicted and actual sensitivities, compare the DMs measured at high and low frequencies, and investigate possible reasons for differing DMs in \S\ref{sec:comparison}.
We measure {a time-dependent power fluctuation known as} baseline ripple seen in pulse profiles obtained with the GMRT, and make predictions for its effect on MSP timing in \S\ref{sec:ripple}. 
We discuss the potential for producing higher precision DMs in \S\ref{sec:inclusion}.

\section{Data}\label{sec:data}

{Observations were done} using the GMRT phased-array mode\footnote{{These data are from the legacy GMRT system, used through 2017. While this has been largely superseded by the upgraded GMRT (uGMRT; \cite{gup17}) some elements of the system are still relevant, and our analysis still holds lessons for future observations. We explicitly compare our results with predictions for uGMRT analysis in \S\ref{sec:comparison}.}}, in which a subset of the full array of antennas can be phased. The phased{-}array mode is capable of off-line coherent dedispersion and allows the array to be sub-divided into two independent {sub-arrays, each of which has its own beam}, allowing for simultaneous dual-frequency observations. 
In order to make the DM measurements at the GMRT useful to aid PTA sensitivity, a dual-frequency coherently-dedispersed observing mode was developed. Significant optimization efforts in computing, memory, and network bandwidth requirements were employed to maintain sustained real-time streaming of dual coherent voltage beams over 32\,MHz bandwidth at Nyquist resolution. These high-gain coherent beams at lower frequencies aided by the high-performance signal processing capability make the GMRT a useful instrument to follow-up PTA MSPs for monitoring ISM parameters. {Earlier work reported by \cite{kum13} for measuring DM variations of PTA MSPs with the GMRT used incoherently dedispersed observing mode.}

Our data were taken simultaneously centered at 322 and 607\,MHz with a 32\,MHz bandwidth at each frequency. A subset of eight antennas was centered at 322\,MHz and 15 antennas centered at 607\,MHz (a maximum of $\sim$23 antennas can be used for phased-array pulsar observations to avoid phasing inefficiency at longer baselines). The observing parameters used here are now considered part of the legacy GMRT system after the implementation of system upgrades to form the upgraded GMRT \citep[uGMRT;][]{gup17}. Predicted sensitivities with the uGMRT are discussed in \S\ref{sec:comparison}. 

\begin{figure}
    \centering
    \includegraphics[trim=1.5cm 1.5cm 1cm 1cm, clip,width=0.47\textwidth]{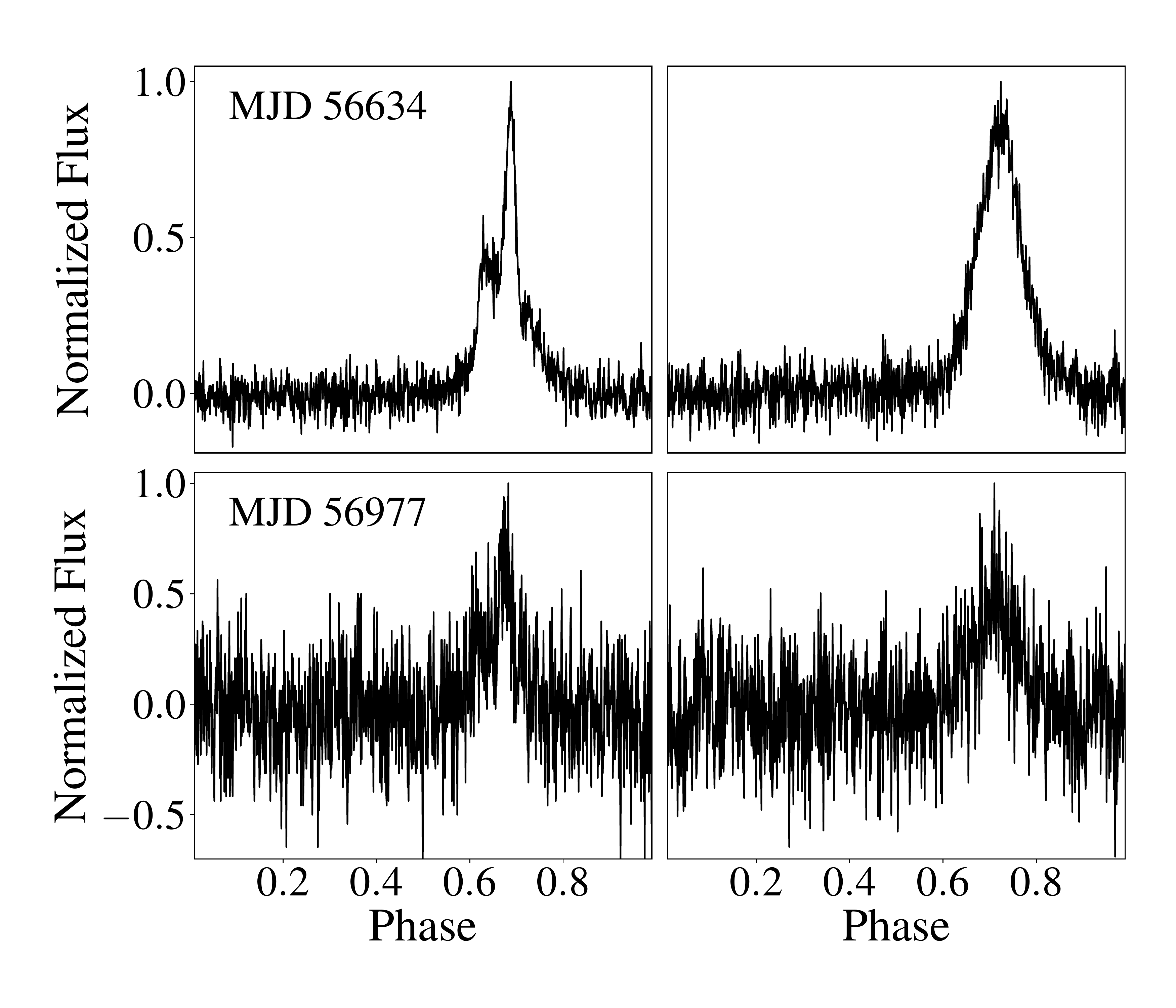}
    \caption{PSR~J1640+2224 pulse profiles for two epochs of GMRT observations at 322\,MHz. The coherently dedispersed data are on the left; the incoherently dedispersed data are on the right. The average {TOA errors of the incoherently dedispersed data are 4 and 10} times higher (on MJDs 56634 and 56977, respectively) when compared to the coherently dedispersed data at 322\,MHz. 
    At 607\,MHz, the difference is predictably less significant with the incoherently dedispersed {TOA errors only double that of the coherently dedispersed data}. }\label{fig:incoherent}
\end{figure}

Observations occurred at 11 epochs between 2013 February 2 and 2014 {October 25}. A test pulsar, PSR B1929+10, was observed at each epoch for $\sim$5 minutes to inspect data quality. 
The GMRT Software Backend simultaneously creates both coherently and incoherently dedispersed data \citep{roy10}; the analysis was performed on the coherently dedispersed pulsar data. DMs for coherent and incoherent dedispersion were obtained from the ATNF Pulsar Catalog \citep[PSRCAT\footnote{http://www.atnf.csiro.au/people/pulsar/psrcat/};][]{man05}. A comparison of the coherently and incoherently dedispersed data can be seen in Fig.~\ref{fig:incoherent}. 

The coherently dedispersed data were split into 32 sub-bands across each of the two frequency bands, while the incoherently dedispersed data was divided into 512 frequency channels. Left and right circular polarizations were combined during processing. 
The data headers were inserted after the observation using a separate script as they are not encoded during the observation. Clock correction files do not exist for the GMRT, and therefore were not used. 

\begin{figure*}
    \centering
    \includegraphics[trim= 1cm 1cm 1cm 1cm,clip, width=0.8\textwidth]{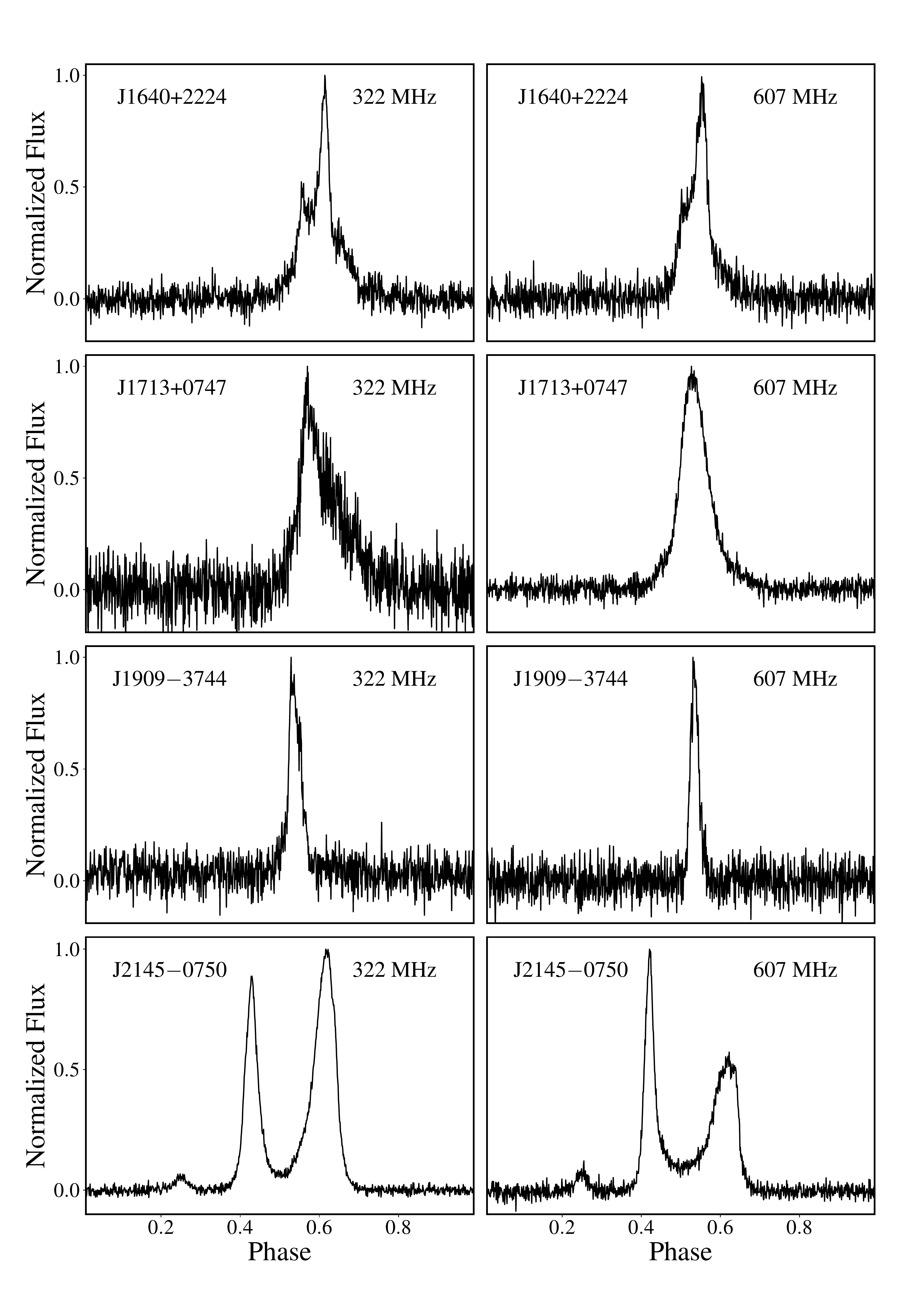}
    \caption{Sample pulse profiles for four MSPs from our GMRT observations.  We show example {single-epoch} pulse profiles at 322\,MHz (left) and 607\,MHz (right) for PSRs~J1640+2224, J1713+0747, J1909$-$3744, and J2145$-$0750 from top to bottom. 
    }
    \label{fig:profiles}
\end{figure*}

Of the ten MSPs originally observed as part of this project, only four had a sufficient number of high signal-to-noise (S/N) detections to be used for high-precision timing. 
PSRs J1640+2224, J1713+0747, J1909$-$3744, and J2145$-$0750 were observed for $\sim$30\,min at each epoch. 
Data were folded using the DSPSR\footnote{http://dspsr.sourceforge.net} \citep{dspsr} software package. 
We fit multiple Gaussians for each pulsar at each frequency for the epoch producing the highest S/N observation to produce a pulse template for calculating TOAs. Examples of pulse profiles from our GMRT observations are shown in Fig.~\ref{fig:profiles}. PSRCHIVE\footnote{http://psrchive.sourceforge.net} was used for TOA generation \citep{van12}. 

\begin{figure}
    \includegraphics[trim = 0.75cm 0.75cm 0cm 0cm, clip,width=0.49\textwidth]{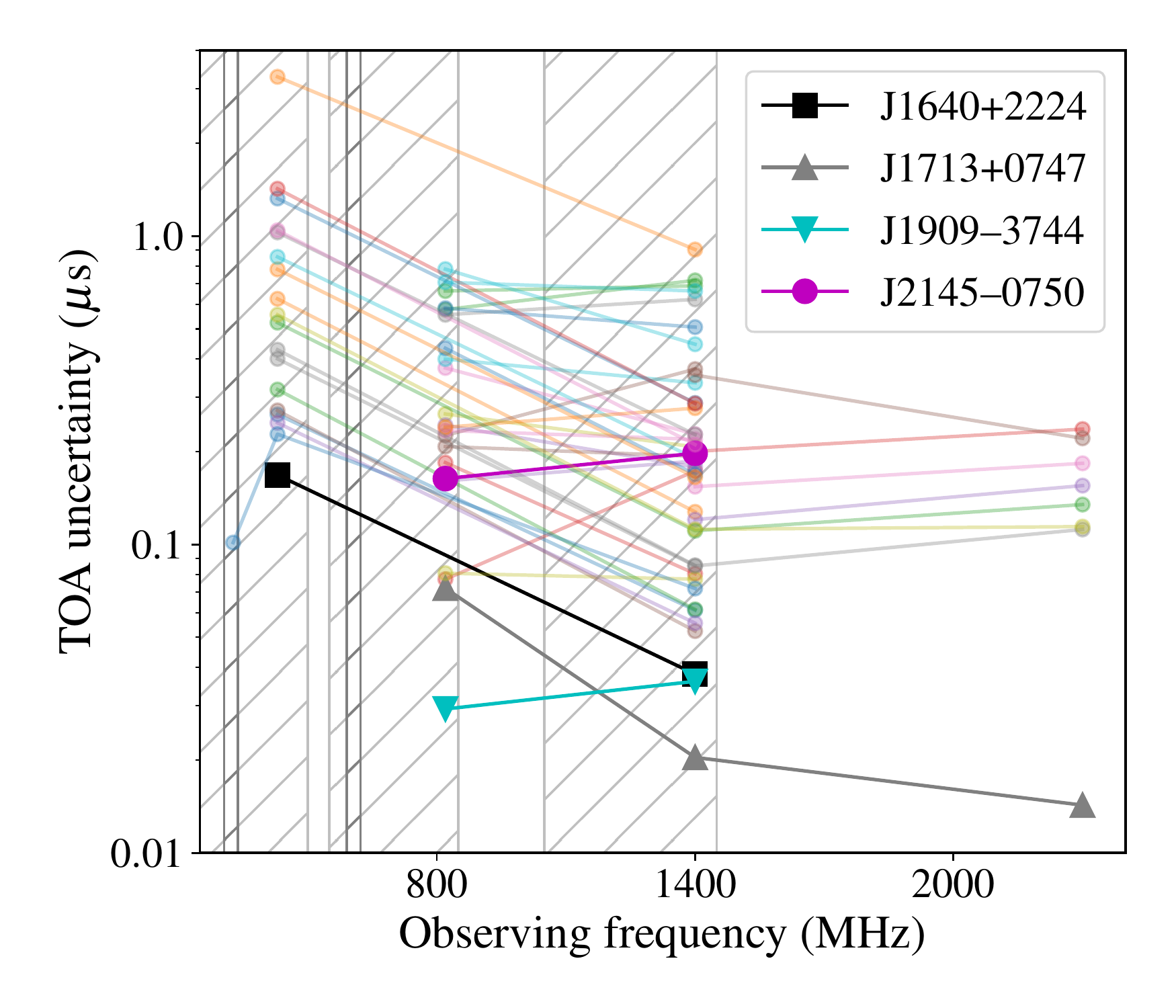}
    \caption{Average TOA uncertainties in the NANOGrav 11-yr dataset  at various center frequencies  \citep{arz18}. The MSPs with GMRT data are the labeled symbols. The lighter gray hatched regions show the potential GMRT  observing frequencies, and the darker gray regions show the two frequencies used in this work. Note that the regions showing the potential observing ranges reflect the current bandwidth capabilities of the GMRT, which have since been upgraded from the 32\,MHz bandwidth available at the time of these observations \citep{gup17}.  
    }
    \label{fig:sources}
\end{figure}

Fitting of the timing model was done using the TEMPO software package\footnote{http://tempo.sourceforge.net}, which applies a least-squares fit to the TOAs \citep{tempo}. When fitting for multiple epochs, the DM values are assumed constant for an individual epoch and encoded via the DMX parameter. 
TEMPO reports 1$\sigma$ errors on DMX   determined from the timing-parameter covariance matrix after the least-squares timing model fit. 
DMX fitting was performed using ephemerides produced from the NANOGrav 11-year dataset \citep{arz18}. The majority of the timing parameters (e.g., positions, binary parameters) from NANOGrav are more precise than those we would obtain from fitting just the GMRT data; therefore all parameters except DMX were held constant in order to obtain a DM estimate {for each epoch} while analyzing the GMRT data. 
Observation frequencies for the NANOGrav data can be seen in Fig.~\ref{fig:sources} for comparison. 
GMRT data could provide low frequency coverage in complement with the NANOGrav data, and fill in the gaps where data below the 820\,MHz band do not exist in the 11-year dataset.

\section{DM prediction and comparison}\label{sec:comparison}

Observing at widely spaced frequencies can decrease DM estimation errors, thus increasing DM precision. 
This more precisely measured DM can then be used to calculate the corrected infinite-frequency TOA. 
Assuming there are no other chromatic timing perturbations, \cite{cor16} show that if we observe at two widely separated frequencies $\nu_1$ and $\nu_2$ {(where $\nu_2 > \nu_1$)} with corresponding average TOA uncertainties $\sigma_1$ and $\sigma_2$, the DM uncertainty can be expressed by
\begin{equation}\label{eq:deltaDM}
    \sigma_{\DM} = \frac{\sqrt{\sigma_1^2+ \sigma_2^2}}{K\left(\nu_1^{-2}- \nu_2^{-2}\right)}~,
\end{equation}
which requires knowledge of the profile alignment across potentially disparate frequency bands. 
When evaluating the estimated DM uncertainty from a single {observing} band measurement centered at $\nu_0$, Eqn.~\ref{eq:deltaDM} {can be approximated by}
\begin{equation}
    \label{eq:single}
\sigma_{\DM} \approx \sqrt{2} \sigmatoa \frac{\nu_0^3}{K \Delta \nu}~,
\end{equation}
where $\Delta \nu$ is the observing bandwidth and $\sigmatoa$ is the average TOA uncertainty for the single band. 
The increase in DM precision with widely spaced frequencies (as opposed to using sub-bands across the bandwidth of one observing frequency) can be estimated for the inclusion of lower frequency data via these two relations. When $\Delta \nu$ is considerably smaller than the difference between frequency bands, it can be seen that the quantity in Eqn.~\ref{eq:deltaDM} is typically much smaller than that in Eqn.~\ref{eq:single}.

However, \cite{cor16} also show that the DM itself is frequency-dependent (chromatic) even at a single epoch due to multi-path scattering; calculating DM at different frequencies will result in different DM estimates because of the net difference in dispersive time delays. This in turn increases the uncertainties in DM estimation over wide bandwidths.

In order to determine the change in DM precision when using multi-frequency observations in comparison with single-frequency observations, we need to examine how the TOA uncertainty changes between bands. 
The mean TOA uncertainty (i.e., averaged over all pulsars) at 1.4\,GHz for the NANOGrav 11-year data is $\sigma_{1.4\,\rm{GHz}}\approx600$\,ns; following Eqn. \ref{eq:single} {for an observation at 1.4\,GHz,} this yields a DM uncertainty of $\sigma_{\rm DM} = 7\times10^{-4}\,\dmunit$ for a typical 800\,MHz bandwidth. 
To achieve the same DM uncertainty through our GMRT observations (where $\Delta \nu = 32$\,MHz),
we would require $\sigma_{322\,\rm{MHz}}=2\,\mu$s and $\sigma_{607\,\rm{MHz}}=0.3\,\mu$s. These uncertainties are smaller than those we were generally able to achieve. However, when {combining the two GMRT bands following Eqn.~ \ref{eq:deltaDM}, the lower limit uncertainties jump to $\sigma_{322\,\rm{MHz}}\approx24\,\mu$s and $\sigma_{607\,\rm{MHz}}\approx3.6\,\mu$s assuming a similar scaling between the two bands as seen between the single TOA uncertainties (which may or may not be the case)}. Adding 1.4\,GHz data, the minimum TOA uncertainty requirements increase even higher to  $\sigma_{322\,\rm{MHz}}=30\,\mu$s and $\sigma_{607\,\rm{MHz}}=7.4\,\mu$s.  

More generally, we can examine how TOA uncertainties change with frequency. We do this for the NANOGrav 11-year data-set (Fig.~\ref{fig:sources}), highlighting the sources with GMRT observations. In general the lower frequency bands have higher TOA uncertainties.   
Typical changes are a factor of $<2$ from 1400\,MHz to 400\,MHz. This would imply that using  Eqn.~\ref{eq:deltaDM} will {result in} significantly {smaller uncertainties in DM, and therefore} more precise DM measurements, from combining the multiple frequencies than either frequency alone.

\subsection{Timing results}

\begin{figure}
    \begin{centering}
    \includegraphics[trim = 0.8cm 1cm 0.75cm 1.5cm,clip,width=0.45\textwidth]{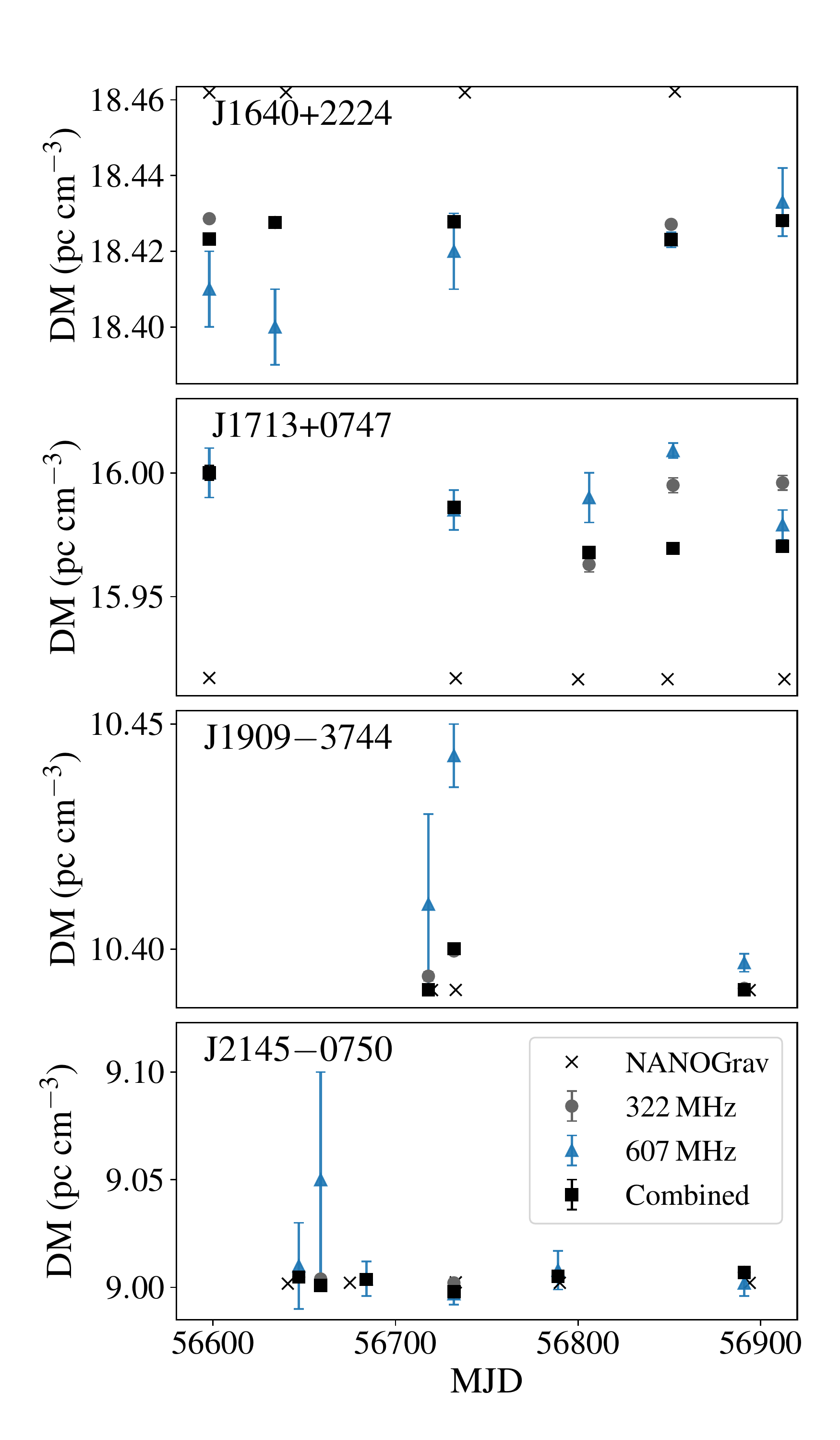}
    \caption{Comparison of {single-epoch} DMs for GMRT and NANOGrav 11-year measurements for the four MSPs considered here: PSRs~J1640+2224, J1713+0747, J1909$-$3744, and J2145$-$0750 form top to bottom. GMRT timing was done using just the 322\,MHz data (circles), just the 607\,MHz data (triangles), and the two frequencies combined (squares). Due to the small variation and small error bars compared to our data, the NANOGrav measurements are contained within the dashed lines.}
    \label{fig:dms}
    \end{centering}
\end{figure} 

{At each epoch,} DMs were measured at 322\,MHz, 607\,MHz, and then fit jointly using both frequencies and assuming no frequency evolution of the pulse profile. 
{The DMs measured from these data are plotted alongside the NANOGrav DMs in Fig.~\ref{fig:dms}}; overall, we see that the GMRT-measured DMs (both single-frequency and combined) have significantly larger uncertainties than the NANOGrav-measured DMs. 
The {Single-epoch} DMs measured here agree with the NANOGrav values for PSR~J2145$-$0750 for all observing epochs and for most epochs for PSR~J1909$-$3744. None of the DMs for J1640+2224 and J1713+0747 agree with the NANOGrav values; the DMs measured for J1640+2224 are consistently much smaller than the NANOGrav DMs by $\sim0.03$\,pc\,cm$^{-3}$, while DMs measured for PSR~J1713+0747 are higher than the NANOGrav values by $\sim0.07$\,pc\,cm$^{-3}$. This is likely a result of unmodeled pulse profile evolution in the two GMRT frequency bands due to the resulting frequency-dependent biases (quantified by FD parameters in the timing model) being held fixed at the NANOGrav values; an independent fit for FD parameters could not be done here due to the small number of observing epochs. A more thorough description of NANOGrav FD parameters can be seen in \cite{arz15}. The average DMs measured across all epochs can be seen in Table~\ref{tbl-timing}. 

\begin{deluxetable}{llclcl}
\tabletypesize{\footnotesize}
%\begin{center}
%\small
\tablecaption{DM estimates from daily timing \label{tbl-timing}}
\tablewidth{0pt}
\tablehead{\multicolumn{1}{c}{PSR}  & \colhead{DM$_{322}$} & $\sigma_{322}$ & DM$_{607}$ & $\sigma_{607}$ & DM$_{\rm{total}}$ \\
& (pc cm$^{-3}$) & ($\mu$s) & (pc cm$^{-3}$) & ($\mu$s) & (pc cm$^{-3}$) 
}
\startdata
J1640+2224  & 18.4281(4) & 1.7 & 18.426(2) & 2.9 & 18.42816(3) \\
J1713+0747  & 15.9888(9) & 8.9 & 16.002(3) & 3.0 & 15.98936(8)\\
J1909$-$3744 & 10.3945(3) & 2.8 & 10.407(2) & 3.6 & 10.459(2) \\
J2145$-$0750 & \phn{9.0042(2)} & 1.5 & \phn{9.012(8)} & 4.2 & \phn{9.00453(3)} 
\enddata
\tablecomments{Results from timing using the GMRT data. Columns list the average DM and average TOA error across all epochs for the 322 and 607\,MHz data respectively, and the DM from timing using both frequencies bands. Quantities in parentheses are 1$\sigma$ uncertainties on the last digit.}
\end{deluxetable}

There are several reasons why DM measurements would differ between the two datasets in addition to pulse profile changes. Non-simultaneous measurements between the different observatories could cause discrepancies. Some of the {single-epoch} DM measurements have $\sim$week-long differences between the NANOGrav and GMRT observing epochs. However, \cite{jon17} calculate the timescales it takes for the DM to vary beyond the measurement errors; the DM variation timescales for the pulsars timed here are all greater than one month, so this is unlikely to be the reason for the discrepancies. 
\cite{lam16} modeled ionospheric DM variations and placed an upper limit to their DM contribution of $\sim 10^{-4}\,\dmunit$, two orders of magnitude smaller than all of the DM differences seen in Fig.~\ref{fig:dms}. 
Following Eqn.~12 in \cite{cor16}, which assumes scattering is due to a thin screen, a fiducial pulsar observed at 322 and 607\,MHz would result in an RMS DM offset due to chromatic DMs of $\sim10^{-4}\,\dmunit$. 
Correcting \textit{only} for DM without correcting for  scattering will cause discrepancies as the DM fit will absorb some scattering effects; however, all four MSPs have DMs below 20\,\dmunit, so they likely do not show sufficient amounts of scattering to be absorbed in the DM modeling. 
Hence, none of these mechanisms are sufficient to explain the scale of the DM offsets we measure.

Since these observations were taken, the GMRT has undergone system improvements to create the uGMRT, including wider observing bandwidth capabilities and more sensitive receiver systems \citep{gup17}. The uGMRT has a maximum instantaneous bandwidth of 400\,MHz; usable bandwidths at the observing frequencies used here are predicted to be 120$-$200\,MHz. {The comparable uGMRT observing bands to those used here are} centered at 400 and 650\,MHz.

{Table~\ref{tbl-uncs} shows the lower limits on the DM precision for each MSP based on the measured TOA uncertainties, as well as the predicted upper limit on TOA uncertainties for future observations required to match the precision of NANOGrav measured DMs.
PSR J2145$-$0750 is the only source timed here for which adding in the GMRT 322\,MHz data in their current state improve the uncertainty on DM. The necessary precision for this MSP could be achieved at 650\,MHz with the uGMRT with a slightly longer ($\sim$40\,min) observation at each epoch using half the array. 
For J1640+2224 and J1909$-$3744, the necessary precision could be reached at 400\,MHz using a similar observing strategy with the uGMRT and, in the case of J1909$-$3744, at 650\,MHz as well; J1640+2224 would require a more robust observing strategy ($\sim$2 hours per epoch with the full array) at 650\,MHz. 
Observations of J1713+0747 could achieve the minimum precision at 400\,MHz by observing for 2 hours per epoch with half of the array, and at 650\,MHz observing for $\sim$2.5 hours per epoch using the full array.
}

\begin{deluxetable*}{c|cc|cl|ll}
\tabletypesize{\footnotesize}
\tablecaption{{GMRT TOA uncertainties and predicted DM precision} }
\tablewidth{0pt}
\tablehead{
\multicolumn{1}{c|}{PSR}  & \colhead{$\sigma_{322}$} & $\sigma_{607}$ & $\sigma_{\rm{DM;GMRT}}$ & $\sigma_{\rm{DM;NG}}$ & $\sigma_{\rm{322;target}}$  & $\sigma_{\rm{607;target}}$ \\
&  ($\mu$s) & ($\mu$s) & (pc cm$^{-3}$) & (pc cm$^{-3}$) & ($\mu$s) &  ($\mu$s)
}
\startdata
J1640+2224  & 0.59--4.3 & 0.75--5.7 & $>1.5\times10^{-5}$ & $9\times10^{-6}$  & $<0.4$ & $<0.07$ \\
J1713+0747  & 3.1--12.7 & 1.1--5.4  & $>8\times10^{-5}$ & $9\times10^{-6}$ & $<0.4$ & $<0.1$    \\
J1909$-$3744 & 0.8--4.5 & 0.7--12 & $>2\times10^{-5}$ & $1\times10^{-5}$ & $<1.2$ & $<0.4$      \\
J2145$-$0750 & 0.71--2.4 & 1.2--12 & $>2\times10^{-5}$ & $5\times10^{-5}$ & $<2.0$ & $<0.4$
\enddata
\tablecomments{{Measured and target GMRT TOA uncertainties. Columns list the range of measured TOA errors across all epochs for the 322 and 607\,MHz data respectively, the lower limit on the GMRT DM uncertainty using the epoch with the lowest TOA uncertainty, the NANOGrav DM uncertainty for that epoch, and the target TOA uncertainties needed to reach the NANOGrav DM precision.}}\label{tbl-uncs}
\end{deluxetable*}

\section{Baseline ripple}\label{sec:ripple}

\begin{figure}
    \centering
    \includegraphics[trim=0.75cm 5.5cm 0.75cm 1.5cm, clip, width=0.48\textwidth]{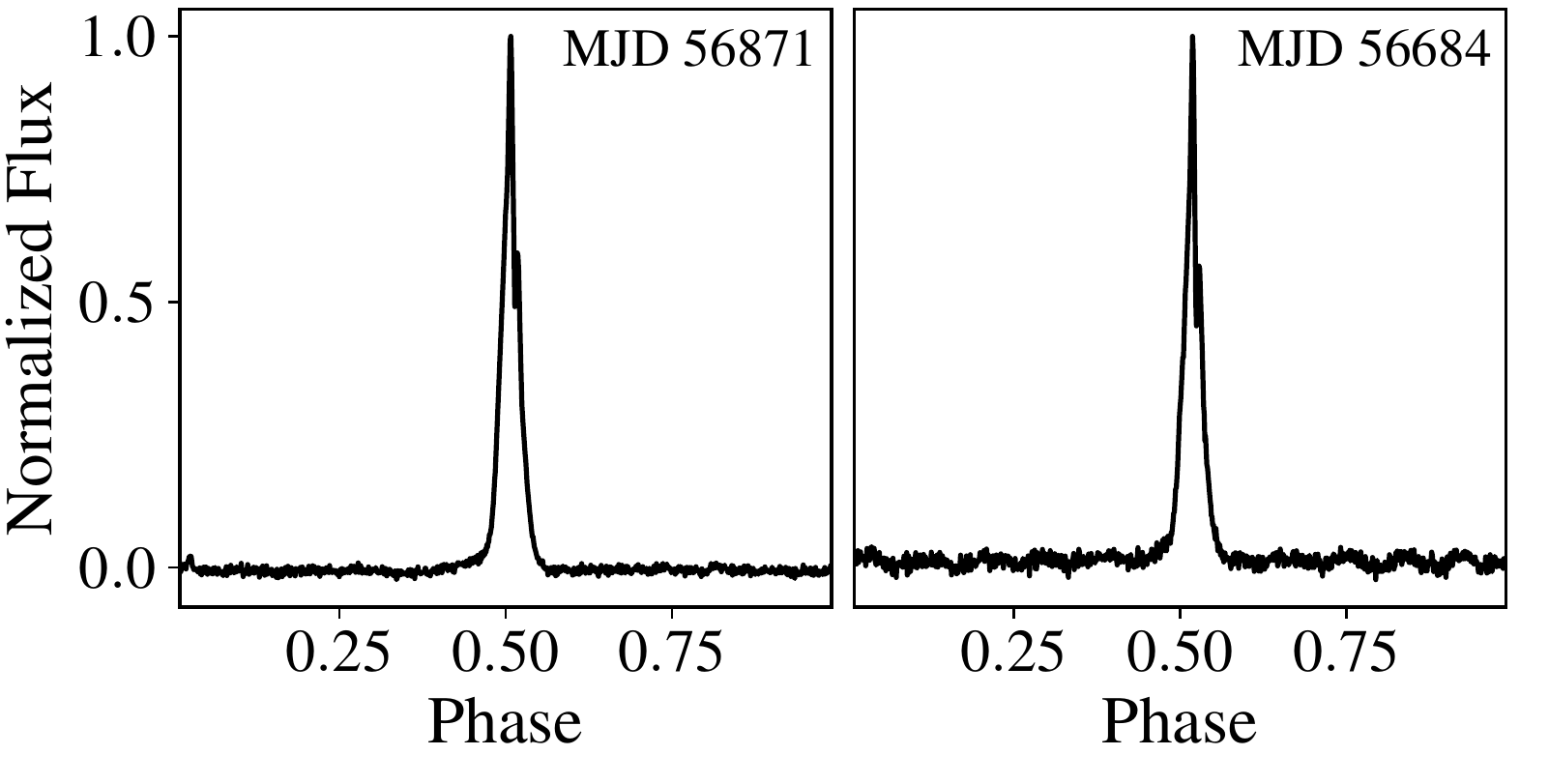}
    \caption{Pulse profile for PSR B1929+10 observed with the GMRT at 607\,MHz on two different epochs separated by about two weeks. The left profile does not show a visible level of baseline ripple, whereas the right profile shows clearly evident ripple in the baseline.}
    \label{fig:ripple}
\end{figure}

Separate from the interstellar medium effects discussed above, additional telescope-specific effects can reduce TOA precision.  In particular, low temporal-frequency (i.e., ``red'') noise in  pulsar profiles can systematically pull a TOA to an earlier or later time, and appears as a stochastic contribution to the TOA error budget. We call this ``baseline ripple", and it could be due to radio-frequency interference, typically due to a nearby power-line or other likely epoch-dependent effects. 
 The phase of the ripple is random relative to the pulse, and therefore is more noticeable for canonical pulsars than MSPs due to the smaller number of times the data are folded over the pulse period (which will be  different than the ripple period). An example of baseline ripple seen in the GMRT data for our test pulsar B1929+10 can be seen in Fig.~\ref{fig:ripple}. While it may not be noticeable by eye for MSPs, it is important that we estimate the effect of baseline ripple on precision MSP timing.

We define a data profile $I(t)$ that is composed of the pulse template $T(t)$ with pulse amplitude $A$ added to a sinusoidal baseline ripple with amplitude $r$, phase $\phi$, and frequency $f_r = \omega_r/2\pi = 1/P_r$,
\begin{equation}\label{eq:template}
    I(t) = A\times T(t-t_0) + r \cos[\omega_r (t-t_0) + \phi]~,
\end{equation} 
where $t_0$ is an arbitrary reference time.
For a Gaussian pulse with full width at half maximum $W$, the template pulse template can be modeled as
\begin{eqnarray}
    T(t) = e^{-4 \ln2 (t/W)^2}~,
\end{eqnarray}
where the approximate error on the TOA becomes
\begin{eqnarray}\label{eq:calcerror}
    \frac{\sigmatoa}{W} = \bigg(\frac{\pi}{4\ln2}\bigg)  \bigg(\frac{r}{A}\bigg) \bigg(\frac{W}{P_r}\bigg) e^{-4\pi\ln2 (W/P_r)^2}~.
\end{eqnarray}
{This derivation is discussed in the appendix.}

\begin{figure}
    \centering
    \includegraphics[trim=0.7cm 0.75cm 0.75cm 0.75cm, clip, width=0.47\textwidth]{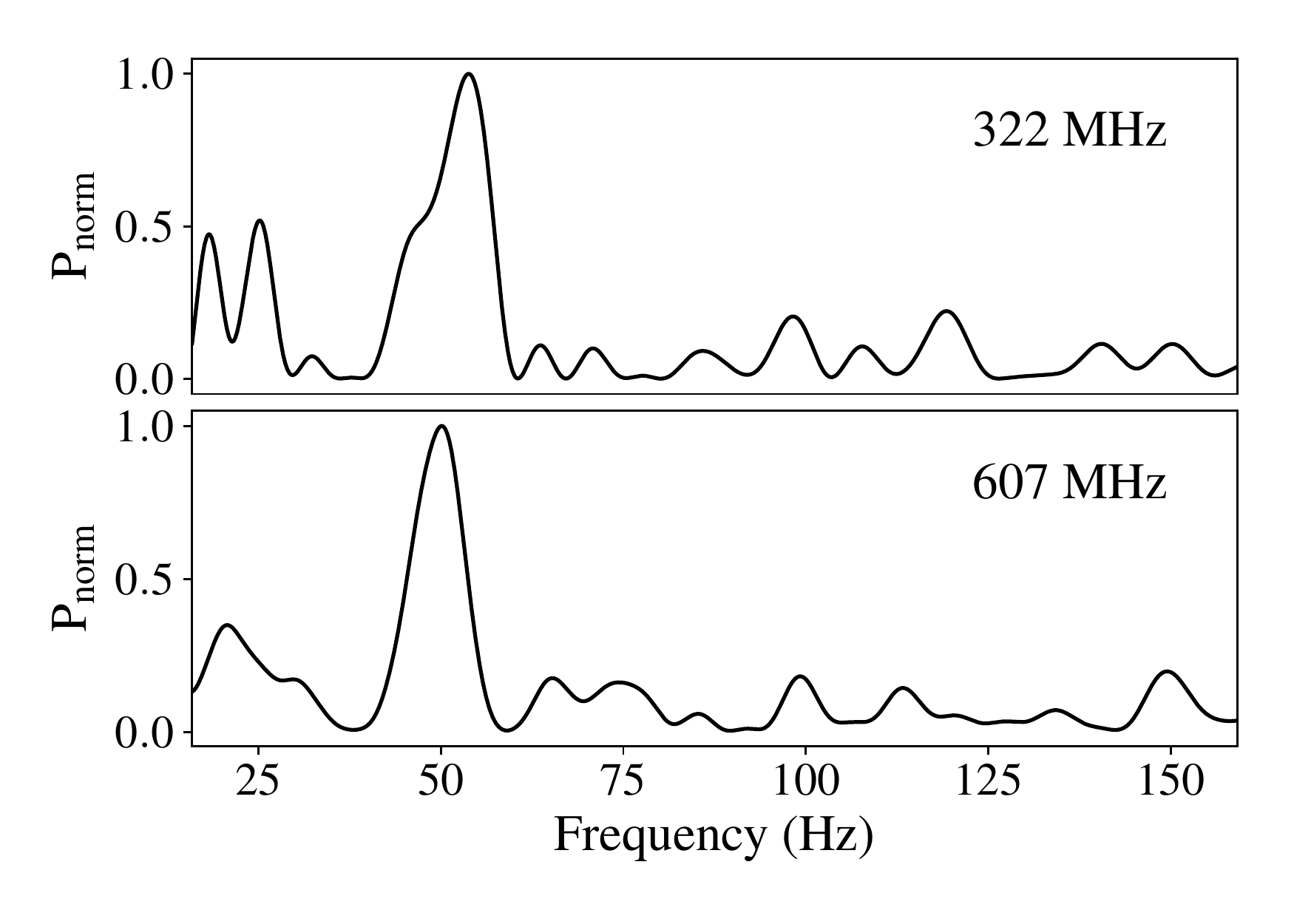}
    \caption{Lomb-Scargle periodograms for the baseline ripple apparent in in PSR B1929+10; the profile can be seen in Fig.~\ref{fig:ripple}. The power is in arbitrary units. The periodogram shows a significant peak ($\sim3-6\sigma$ depending on the epoch) at approximately 52\,Hz. Epochs that showed peaks below 3$\sigma$ significance were not included in this analysis.}
    \label{fig:scargle}
\end{figure}

{Because the ripple and pulse periods are likely non-harmonic, we use a Lomb-Scargle periodogram to search for any latent periodicity.} Applying a Lomb-Scargle periodogram to the pulse profile for our test pulsar PSR B1929+10, we detect a ripple frequency of $f_r = 52\pm5$\,Hz, seen in Fig.~\ref{fig:scargle}. {This matches with the AC power-line frequency in India (50\,Hz).} Epochs where the detected ripple was below a $3\sigma$ significance threshold were not included in this estimate; only one epoch did not show a detectable ripple ($<1\sigma$ peak). 
The induced timing errors due to ripple for the four MSPs can be estimated via Eqn.~\ref{eq:calcerror} using our observations of PSR B1929+10 to measure the amplitude $r$ of the baseline ripple. Scaling the detected ripple amplitude to the MSP flux densities relative to B1929+10 at each epoch, we estimate $r/A\approx
0.03$ in our MSP observations. With the respective pulse widths and periods, this signal can cause timing uncertainties up to $\sim$150\,ns for J1640+2224 and J1909$-$3744, up to $\sim$340\,ns for J1713+0747, and up to a microsecond for J2145$-$0750 (due to its longer period and larger pulse width). 
Given that the achievable GMRT timing precisions in this work are a few microseconds or greater, baseline ripple does not appear to be a concern for these data (but a similar signal could be important for NANOGrav data taken with the GBT). This effect will need to be considered for high-precision pulsar timing observations with the uGMRT.

\section{Interstellar scintillation}\label{sec:scintillation}

{In addition to timing analyses, measurements of the modulated radio pulsar emission due to scintillation in the ISM can directly probe the properties of the intervening material. \ch{Scintillation bandwidths can be used to estimate delays from pulse broadening on the TOAs, and can therefore help to track scattering delays and discriminate between those and DM delays.} Scintillation observations do not require the timing accuracy or precision of pulsar timing experiments. Measurements of scintles, intensity maxima in a dynamic spectrum $I(t, \nu)$, require high S/N pulse profiles observed over sufficiently large bandwidths to capture the maxima, but also high frequency and time resolution to discriminate individual scintles. The characteristic bandwidth and timescale of scintles vary as different functions of frequency, with both becoming smaller at lower radio frequencies. GMRT observations at low frequencies can therefore extract useful information about ISM, in particular for the lowest DM pulsars with the largest scintle sizes.
}

\begin{figure}
    \centering
    \includegraphics[width=0.47\textwidth]{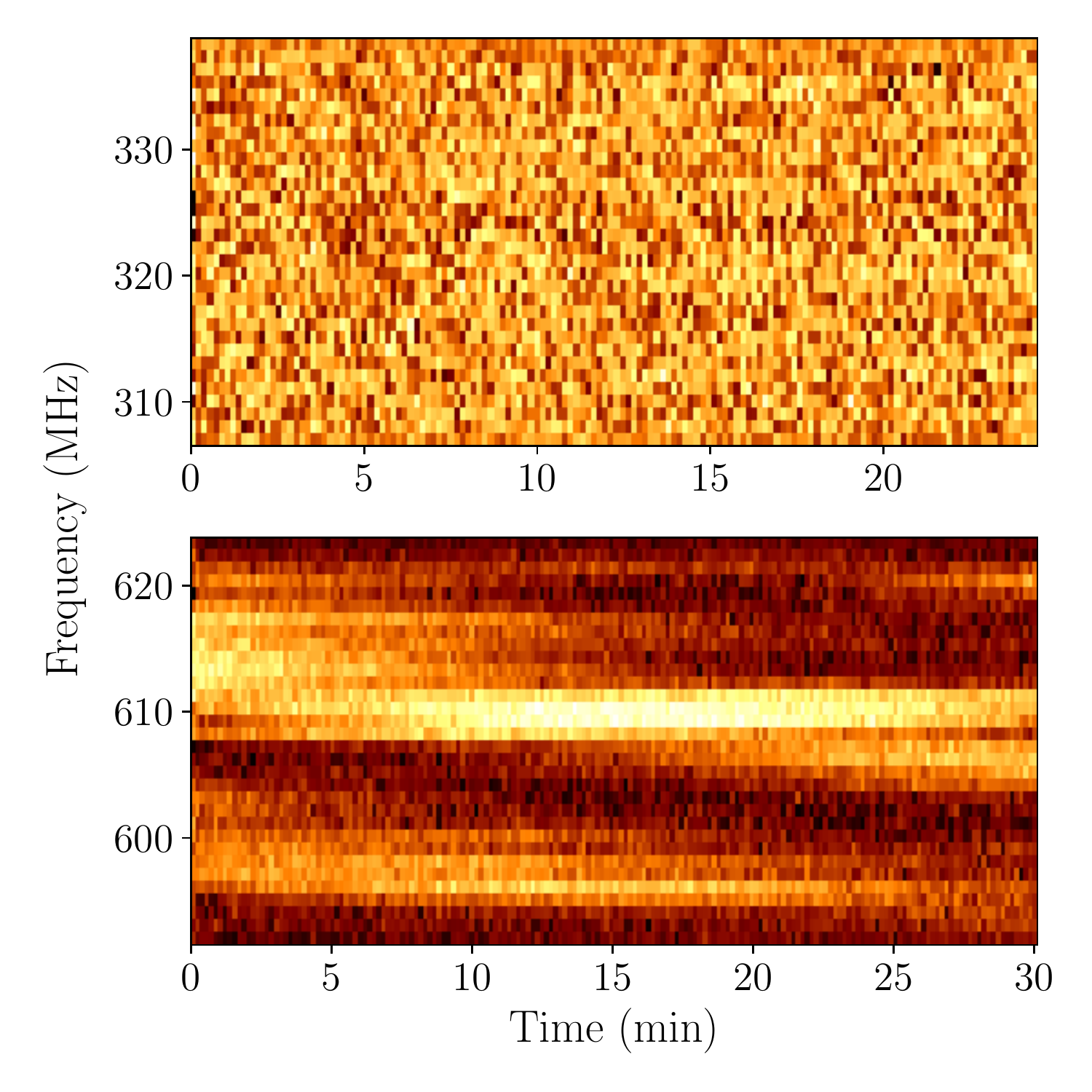}
    \caption{{Dynamic spectra for PSR J1713+0747 on MJD 56852. The scintles can be clearly seen in the 600\,MHz band. Note that the {total time spanned by the observation} may vary by several minutes between the two bands.}}
    \label{fig:ds_1713}
\end{figure}

\begin{figure}
    \centering
    \includegraphics[width=0.47\textwidth]{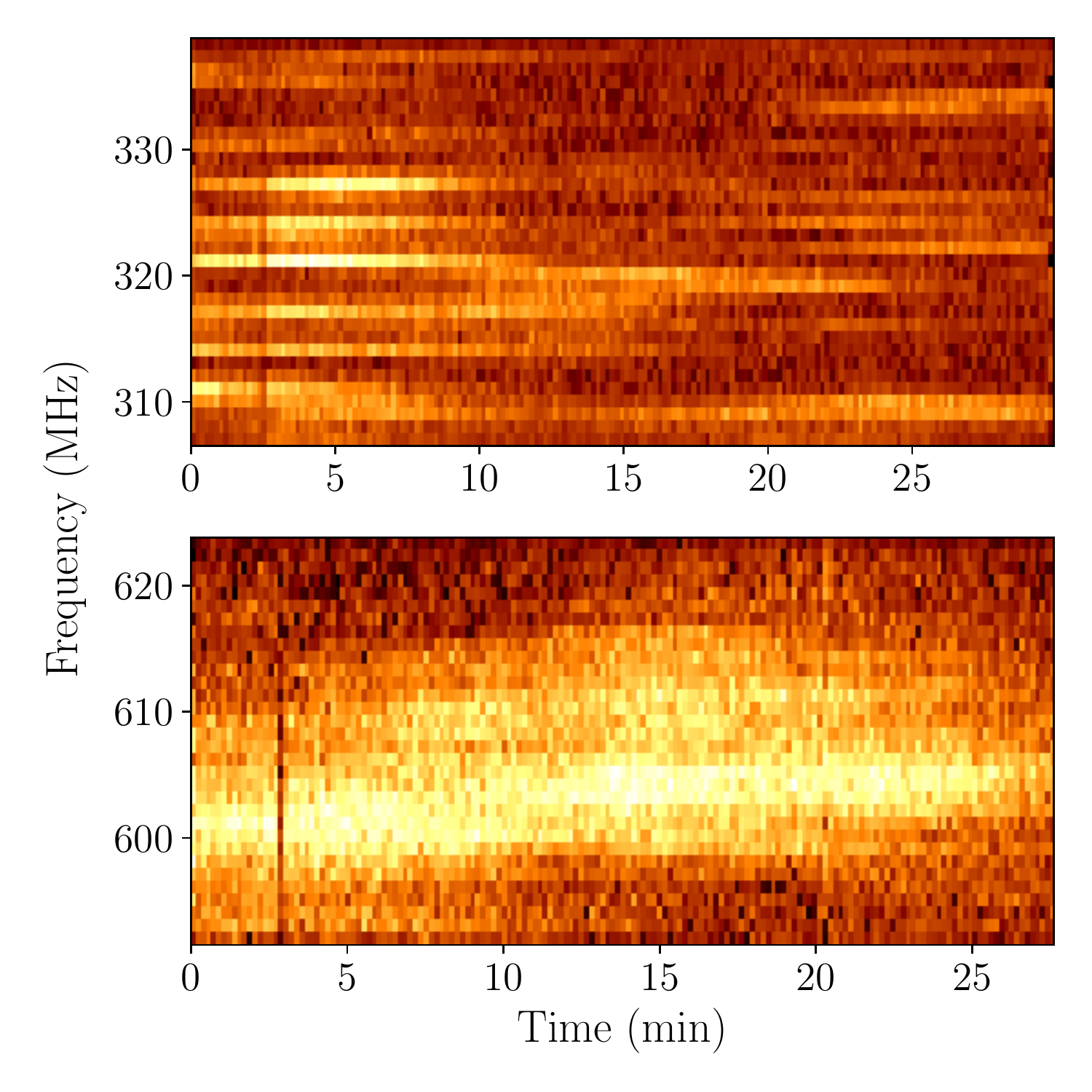}
    \caption{{Dynamic spectra for PSR J2145$-$0750 on MJD 56732. The scintles can be clearly seen in both the 300 and 600\,MHz bands. Note that the {total time spanned by the observation} may vary by several minutes between the two bands.}}
    \label{fig:ds_2145}
\end{figure}

{The standard procedure for measuring the scintillation parameters, the characteristic bandwidth $\Delta \nu_{\rm d}$ and timescale $\Delta t_{\rm d}$, from the dynamic spectrum is via an autocorrelation function (ACF). For a dynamic spectrum $I(t, \nu)$, we define the 2D ACF as
\be 
R_I(\delta t, \delta \nu) = \langle I(t, \nu) I(t + \delta t, \nu + \delta \nu) \rangle ~.
\ee
In the case when the S/N values of ACFs determined for each epoch are low, we can take the ACFs and add them together to increase the total S/N with which to measure scintillation parameters:  
    \be 
    R_I (\delta t, \delta \nu) = \sum_n R_{I,n}(\delta t, \delta \nu),
    \ee
where $n$ is the index over the epoch.  This procedure assumes that variations in the scintillation parameters are small relative to the uncertainties. {While single-epoch ACFs  typically had low S/N, we were able to measure  average scintillation parameters across all datasets  with an increased S/N}. Using \textsc{PyPulse}\footnote{https://github.com/mtlam/PyPulse} \citep{pypulse}, we generated ACFs for epochs when distinct scintles were seen. ACFs were added together based on the appearance of scintles in the band, which enabled more robust measurements of the scintillation bandwidths. Dynamic spectra produced for PSRs J1713+0747 and J2145$-$0750 can be seen in Figures \ref{fig:ds_1713} and \ref{fig:ds_2145}.} 

{For PSR~J1713+0747, we co-added ACFs from five epochs of the 607\,MHz data (56806, 56852, 56912, 56598, and 56955), again when distinct scintles were seen. For PSR~J2145$-$0750, we co-added ACFs from two epochs of 322\,MHz data (56732 and 56891). We fit a 1D Gaussian to the slice of the 2D ACF at zero time lag, i.e., $R_I(0, \delta\nu)$, and calculated the half-width at half maximum as is standard \citep[e.g.,][]{cor02b}. For PSR~J1713+0747, we measured $\Delta \nu_{\rm d, 607} = 3.1$\,MHz, and for PSR~J2145--0750 $\Delta \nu_{\rm d, 322} = 3.6$\,MHz. Scintillation bandwidth scales with observing frequency as 
\be
 \Delta\nu_{\rm{d}} = \Delta\nu_0 \left(\frac{\nu}{\nu_0}\right)^{\xi}~,
 \ee
where $\nu$ and $\nu_0$ are the higher and lower observing frequencies, $\Delta\nu_{\rm{d}}$ and $\Delta\nu_0$ are the higher and lower scintillation bandwidths, and $\xi$ is the scaling factor. Analysis of the NANOGrav 11-year data set found $\Delta \nu_{\rm d, 1400} = 21.1\pm 8.6$\,MHz and $47.8\pm 13.3$\,MHz at $\nu = 1400$\,MHz for J1713+0747 and J2145$-$0750 respectively \citep{lev16}. Comparing these two measurements corresponds to a scaling factor of $\xi=2.3\pm0.5$ for J1713+0747 and $\xi=1.8\pm0.2$ for J2145$-$0750. 
The scintillation bandwidth is expected to scale in frequency as $\nu^{22/5}$ for a Kolmogorov medium with a single thin scattering screen \citep{cor02}. }
{Frequency-dependent scintillation bandwidth scaling shallower than 22/5 is not unexpected and has been seen in other analyses \citep[e.g.,][]{bha04,lev16,tur20}.} 
{A possible explanation for the discrepancy between the measurements is that there are two scintillation scales. Finding multiple scales ({within an order of magnitude}) of scintillation is a known effect for nearby pulsars \citep{gwi06}. Longer-term monitoring of the scintillation parameters at lower frequencies will be required to make more definitive claims regarding these discrepancies. Note that we should treat {these single-epoch scintillation} measurements with caution, as it is well known that the scintillation bandwidth can vary dramatically from epoch to epoch \citep{col15}.}

{The scintillation bandwidth can be used to calculated the scattering delay $\tau_{\rm d} = C_1/2\pi\Delta\nu_{\rm d}$, where $C_1$ is a constant that varies with the geometry and spectral model of the ISM; we adopt $C_1=0.654$ which corresponds to a Kolmogorov medium with a thin scattering screen \citep{lam99}. This yields a scattering delay of $\tau_{\rm d} \approx$ 30\,ns for both pulsars. These delays are $\sim$2 orders of magnitude smaller than the TOA errors we measure here and therefore are not an issue in these data. However scattering correction may become important for low-frequency observations used for GW detection where the goal is $\lesssim$100\,ns timing precision.}

{The transverse velocity of the pulsar can be estimated using the scintillation timescale 
\be
    V_{\rm ISS} = A_{\rm ISS} \frac{\sqrt{\Delta\nu_{\rm{d}}D_{\rm{kpc}}x}}{\nu_{\rm{GHz}}\Delta\tau_{\rm{d}}}~,
\ee
where $A_{\rm{ISS}}=2.53\times 10^4$\,km/s for a Kolmogorov medium and $x=D_o/D_p${, where $D_o$ is the distance between the screen and the observer and $D_p$ is the distance between the screen and the pulsar} \citep{gup94,cor98,tur20}; we assume $x=1$ for a screen halfway along the LOS. Here we use a lower limit equal to the length of our observations $\Delta\tau_{\rm{d}} \geq $30\,min, which was not long enough to characterize the scintillation timescale. 
We can therefore calculate upper limits of the transverse velocity of  $V_{\rm{ISS}} < 43$\,km/s and $V_{\rm{ISS}} < 74$\,km/s for for J1713+0747 and J2145$-$0750, respectively; these agree with published velocities derived through proper motion measurements \citep{tur20}.}

\section{Conclusions}\label{sec:inclusion}

This work illustrates the {one of the} first attempt{s} to utilize the GMRT for IPTA work. 
We have demonstrated that the GMRT can be successfully used to time MSPs and measure DM, in some cases with comparable sensitivity to current PTA data. We observed with only a subset of the array for the data presented here; one factor that would go a long way in improving data precision is using the complete array for timing observations. {With the availability of larger instantaneous bandwidth (e.g. 300--500\,MHz) for the uGMRT, the similar observations for monitoring the ISM effects can be carried out with the complete array.}
The GMRT observer's manual\footnote{http://www.gmrt.ncra.tifr.res.in/} predicts the array gain as $\sim$0.33\,K\,Jy$^{-1}$ antenna$^{-1}$; if the entire array were used (double the maximum number of antennas used here), the predicted gain increases to $\sim$10\,K\,Jy$^{-1}$, assuming no losses due to beamforming. For comparison, the Arecibo Observatory lists a gain of 11\,K\,Jy$^{-1}$ at similar frequencies. 

For all of the MSPs discussed above, known DM effects could not account for the offset seen between {single-epoch} DM measurements, which suggests that the sources of these variations are not due to the ISM. 
Due to chromatic DMs, we would not expect agreement between DMs measured at different frequencies, but as discussed above {the measured differences are larger than can be explained by chromatic DMs}.  
DMs may show variability due to a combination of scintillation and pulse profile evolution, as we essentially see a different part of the pulse at each epoch.  
The NANOGrav dataset includes FD parameters in the timing model, which account for pulse profile evolution, while this analysis did not due to the limited number of observing epochs. This combination is the likely source of a DM offset between frequencies. \footnote{{After the submission of this work, the uGMRT has seen similar DM offsets to NANOGrav DMs; these offsets are also attributed to variations in pulse profile templates \citep{kri21}}}

As mentioned in Section \ref{sec:data}, polarization calibration was not done when the GMRT data used here were obtained. Not accounting for polarization causes TOA uncertainties due to deviations from the pulse profile template. Using fiducial values for NANOGrav data, the anticipated TOA uncertainty induced by errors in polarization calibration to be $\sim100~$ns$-1~\mu$s \citep{van06,lam18} for narrow frequency channels (this value averages down {when integrating in time}, but may change systematically between epochs); as no polarization calibration was performed, the errors for the GMRT data are likely larger. {Without significant standardizations in high-fidelity polarization calibration, TOAs obtained by the uGMRT system will be difficult to integrate into global PTA efforts. }

While the data presented here overall do not meet the required TOA uncertainties for an improved DM measurement, high-precision timing required by PTAs appears feasible with the upgraded capabilities of the uGMRT.  
This sensitivity should be achievable with a similar observing strategy used here, and in some cases with longer observation times and larger subsets of the array. Even without PTA-level timing precision, the lower frequency timing data of the GMRT will still provide valuable science related to the ISM and propagation effects. {Lower-frequency data show smaller scintles than at higher frequencies; the increased bandwidth of the uGMRT systems will effectively capture more scintles and greatly boost the S/N of ACF measurements. Given the smaller scintle sizes, the uGMRT should prioritize short subintegrations and small frequency-channel resolutions in their observations. While costly in terms of data volume, it is possible to develop real-time pipelines to save these data products. Measurements of dynamic spectra over long timespans are critical in helping to constrain the properties of the turbulent ionized ISM. In turn, even without high-precision TOAs, these constraints can feed into PTA analyses, allowing for improved mitigation of scattering effects among all pulsars in the array. 
{The high-precision DMs provided by the uGMRT could play a very important role in IPTA datasets going forward, especially giving the recent loss of the Arecibo telescope.
}

\acknowledgments
The NANOGrav collaboration is supported by NSF Physics Frontier Center award \#1430284. MTL also acknowledges support from NSF AAG award number \#2009468. 
We acknowledge support of the Department of Atomic Energy, Government of India, under project no. 12-R\&D-TFR-5.02-0700. 
The GMRT is run by the National Centre for Radio Astrophysics of the Tata Institute of Fundamental Research, India. We 
acknowledge support of GMRT telescope operators for observations.

\software{DSPSR \citep{dspsr}, PSRCHIVE \citep{psrchive}, PyPulse \citep{pypulse}, TEMPO \citep{tempo}}

\facility{GMRT}
\clearpage

\begin{appendix}

\section{TOA Offsets due to Baseline Ripple}\label{sec-ripple}
\setcounter{equation}{0}

{Here we derive the perturbations to pulse arrival times due to an unmodeled sinusoidal ripple. We define a data profile $I(t)$ that is composed of the pulse template $\U(t)$ with amplitude $A$ {and arrival time $t_0$}, and some sinusoidal baseline ripple with amplitude $r$, phase $\phi$, and temporal frequency $f = \omega/2\pi = 1/P_r$,}
\begin{equation}
    I(t) = A\times \U(t-t_0) + r \cos[\omega (t-t_0) + \phi]~.
\label{eqn:ripple}
\end{equation}
{The template fitting procedure used to calculate the TOA can be represented in terms of finding the maximum cross-correlation of the data $I$ and template $\U$, $C_{I\U} (\tau)$, the solution of which gives the TOA $\hat{\tau}$:}
\begin{eqnarray}
    C_{I\U} (\tau) = \int dt I(t) \U(t-\tau) \\ 
    \Longrightarrow \frac{dC_{I\U}}{d\tau} (\hat{\tau}) = 0 ~.
\label{eqn:dCdtau}
\end{eqnarray}
{We can expand around $\U'(t-\tau)$ about $\U'(t-t_0)$ to first order by assuming that the TOA error $\delta \tau= \hat{\tau}-t_0$ is much smaller that the template pulse width $W$. Doing this expansion and solving for the TOA error in Eqn.~\ref{eqn:dCdtau} gives}
\begin{eqnarray}
    \delta \tau = \hat{\tau} - t_0 \approx \frac{\int dt I(t) \U'(t)}{\int dt I(t) \U''(t)}~.   
    \label{eqn:dtau}
\end{eqnarray}
{Plugging in Eqn.~\ref{eqn:ripple} yields}
\begin{eqnarray}
    \delta \tau = - \frac{(rf/2\pi) \Im \left\{e^{-i\phi} \tilde{\U}(f)\right\}}{A \int df' f'^2 |\tilde{\U}(f')|^2 + f^2 r \Re\left\{ e^{-i\phi} \tilde{\U}(f) \right\}}~,
\end{eqnarray}
{where $\tilde{\U}(f)$ if the Fourier transform of the template using $e^{-2\pi ift}$ and $\Re$ and $\Im$ are the real and imaginary parts respectively.}

{For $r/A \ll 1$,}
\begin{eqnarray}
    \delta \tau = - \bigg( \frac{rf}{2\pi A} \bigg) \frac{\Im \left\{e^{-i\phi} \tilde{\U}(f)\right\}}{\int df'  f'^2|\tilde{\U}(f')|^2} ~.
\label{eqn:dtausmall}
\end{eqnarray}
{Multiple ripple terms then add linearly to the net TOA error; multiple sinusoids can be considered as a Fourier sum.}

{For a Gaussian pulse with full width at half maximum $W$,}
\begin{eqnarray}
    \U(t) = e^{-4 \ln2 (t/W)^2}~,
\end{eqnarray}
{the approximate error given by Eqn.~\ref{eqn:dtausmall} becomes}
\begin{eqnarray}
    \frac{\delta \tau}{W} = \bigg(\frac{\pi}{4\ln2}\bigg)  \bigg(\frac{r}{A}\bigg) \bigg(\frac{W}{P_r}\bigg) e^{-\pi^2 (W/P_r)^2/4\ln2} \sin\phi~,
\label{eqn:dtaugauss}
\end{eqnarray}
{which is used to estimate the ripple-induced TOA error in \S \ref{sec:ripple} (eqn. \ref{eq:calcerror})}
{The maximum of $(W/P_r)e^{-\pi(W/P_r)^2/4\ln 2}$ occurs at $W/P_r= \sqrt{2\ln2}/\pi = 0.375$, corresponding to a maximum error}
\begin{eqnarray}
    \bigg( \frac{\delta \tau / W}{r (\sin\phi)/A} \bigg) = (8e \ln2)^{-1/2} = 0.258~.
\end{eqnarray}
{For a multi-epoch dataset, we expect $\phi$ to be uniformly distributed in $[0, 2\pi]$ yielding $\sigma_{\sin \phi}= 1/\sqrt{2}$. Then for $r = {\rm constant}$ at all epochs, the maximum RMS error is
}\begin{equation}
    \left(\frac{\sigma_{\delta \tau}/W}{r/A}\right)_{\rm max} = \sigma_{\rm \sin \phi} \left(8e \ln 2\right)^{-1/2} = \left(4\sqrt{e\ln 2}\right)^{-1}=0.182~.
\end{equation}

{
\section{Mean amplitude of folded ripple sinusoid}
Here we derive how the ripple amplitude changes, and therefore induced TOA offsets, when folding the ripple sinusoid over different pulse periods. We define a sinusoidal baseline ripple $x(t) = r\sin (\omega t)$. The time $t$ is related to the pulsar phase as $t=m (P +\delta \theta$), where $m$ is an integer and the fractional phase $\delta \theta$ is between 0 and 1.
The ripple evaluated at the pulse phase is 
\be
    x(\delta\theta)_m =  r\sin[\omega (m P + \delta\theta P)]~.
\ee
The profile of the baseline ripple averaged over a number of $M$ folded pulses  is given by
\be
    S(\delta\theta) = \frac{1}{M} \sum_{m=0}^{M-1} x(\delta\theta)_m = \frac{1}{M} \sum_{m=0}^{M-1} re^{i(\omega m P + \omega\delta\theta P)}~.
\ee
Evaluating the sum shows how the ripple profile varies sinusoidally in the pulse phase
\be
    S(\delta\theta) = e^{i\delta\theta P}  \left(\frac{r\sin (\omega M P/2)}{M\sin (\omega P/2)}\right)~.
\ee
The term in parentheses is the amplitude of the ripple, so we can write:
\be
|S| = \left|\frac{r\sin (\omega M P/2)}{M\sin (\omega P/2)}\right|.
\ee
As $\omega = 2\pi/P_r$, we let the ratio of the spin period and ripple period $P/P_r = k+\delta\psi$ where $k$ is an integer (this assumes $P>P_r$, but this can easily be adapted to $P<P_r$). The ripple amplitude then becomes
\be
    |S| = \left| \frac{r \sin (\pi M \delta\psi)}{M\sin (\pi\delta\psi)} \right|~.
\ee
When the two periods are harmonically related ($\delta\psi=0$), the ripple amplitude does not decrease with folding. Otherwise the exact ripple amplitude depends on $\delta \psi$ (and is highest for $\delta \psi <1/M$) but the expected value decreases as $1/M$.  
}
\end{appendix}

\bibliography{gmrt}

\end{document}